\documentclass{article}
\usepackage{graphicx}
\title{Random Substitution-Insertion-Deletion (RSID) Model of Molecular Evolution with Alignment-free Parameter Estimation}
\author{David Koslicki}
\usepackage{graphicx}
\usepackage{amsmath,amsthm,amsfonts,latexsym,amscd}
\usepackage{bbm}
\usepackage[mathscr]{eucal}
\usepackage{amssymb}
\usepackage{wrapfig}
\newtheorem{lemma}{Lemma}[section]

\newtheorem{definition}{Definition}[section]

\begin{document}
%\twocolumn

\maketitle

\begin{abstract}
We present a comprehensive new framework for handling biologically accurate models of molecular evolution. This model provides a systematic framework for studying models of molecular evolution that implement heterogeneous rates, conservation of reading frame, differing rates of insertion and deletion, customizable parametrization of the probabilities and types of substitutions, insertions, and deletions, as well as neighboring dependencies. We have stated the model in terms of an infinite state Markov chain in order to maximize the number of applicable theorems useful in the analysis of the model. We use such theorems to develop an alignment-free parameter estimation technique. This alignment-free technique circumvents many of the nuanced issues related to alignment-dependent estimation. We then apply an implementation of our model to reproduce (in a completely alignment-free fashion) some observed results of Zhang and Gerstein \cite{Zhang2003} regarding indel length distribution in human ribosomal protein pseudogenes.
\end{abstract}

\section{Introduction}
%Accurate models of molecular evolution are indispensable for the proper analysis of biological processes. In particular, phylogenetic inference from sequence data relies heavily upon the chosen model of molecular evolution and the particular parameter values that have been used (***cite***). Traditional stochastic models for DNA or amino acid sequence evolution are based upon 4 or 20 state Markov processes. For example, in the DNA case, the Jukes and Cantor (***cite***), Kimura two parameter(***cite***), Felsenstein (***cite***), HKY (***cite***), and Reversible (***cite***) models of molecular evolution all defined via the specification of the 12 parameters in the instantaneous rate matrix (***cite lio and goldman***). While these models have been referred to as ``state-of-the-art'' (**cite whelan**), none of these oft-used models incorporate insertions and deletions (indels). Attempts to implement indels in the past have been made by Thorne, Kishino, and Felsenstein (**cite**) but non-biological entities (referred to as ``fragments'') whose boundaries do not change over time where used to simplify the pertinent calculations. Little work has since been done in the way of presenting a comprehensive mathematical environment in which biologically accurate models of molecular evolution can be studied.
While substitution models of molecular evolution have been well studies and developed, the inclusion of insertions and deletions (indels) into biologically accurate models has enjoyed less success. As remarked in \cite{Bradley2007}, a robust and comprehensive understanding of probabilistic indel analysis (and its relation to sequence alignment) is still lacking. A number of models of molecular evolution that include substitutions, insertions and deletions have been proposed (\cite{Cartwright2009}, \cite{Miklos2009}, \cite{Miklos2004}, \cite{Thorne1991}, \cite{Thorne1992}), but what is missing is a comprehensive mathematical structure in which can be cast the problem developing of biologically accurate models of molecular evolution. In fact, it is this lack of a well-studied mathematical structure that leads to the analytic intractability of some proposed indel models (as mentioned in \cite{Miklos2004}). This lack of a unifying structure not only gives rise to computational entities with no biological counterpart such as ``fragments'' (\cite{Thorne1991}, \cite{Thorne1992}) and the embedding of a given sequence into an infinite sequence (\cite{Miklos2004}), but also leads to difficulties in comparing models, their assumptions, and their applicability. For example, the relationships between various substitution models of molecular evolution are well understood and these relationships can be easily compared and contrasted (as in \cite{Whelan2001} and \cite{Lio1998}). This is due to most traditional substitution models being stated in terms of instantaneous rate matrices of a finite state Markov chain. Due to the variety of mathematical tools that have been used to implement indels in various applications (for example: HMM's \cite{Cartwright2009}, rate grammars \cite{Miklos2004}, transducers \cite{Kim2007}, birth-death processes \cite{Thorne1991} \cite{Thorne1992}) no such immediate comparison of models is possible.

A few structures have been suggested, most notably the framework of a finite state transducer (\cite{Bradley2007}). While finite state transducers indeed seem promising, we take a slightly different route that comes from the direction of dynamical systems (and so will allow for the application of many well-known dynamical systems' tools).

% they have yet to be successfully implemented to provide a robust model of molecular evolution that, upon being given a set of \textit{unaligned sequences}, produces evolutionarily relevant statistics like substitution rate, indel length distribution, limiting codon frequency distribution, etc.

We present in this paper a comprehensive mathematical theory in which one can explore biologically accurate models of molecular evolution. We also present an alignment-free method of estimating parameters.

Our model allows for the incorporation of substitutions, insertions and deletions of any length less than or equal to a specified length $N$. We also incorporate rate heterogeneity, and neighboring dependencies (context dependency) up to a specified distance. The model is discrete time (it can be viewed as a generalization of a stochastic grammar) and the assumptions are clearly stated: First, we assume that in one time unit, (possibly different) segments of DNA will not be both deleted and inserted. In one step, only insertions or deletions are allowed, not both. Second, we assume reversibility, though as we will see, this assumption can easily be relaxed. These are the only inherent assumptions in this model, but since our model contains a high degree of flexibility further assumptions can be made if, for example, one insists on using, say, the HKY (\cite{Hasegawa1985}) model as the underlying substitution model.

The mathematical language in which we cast this model is that of a discrete, time-homogeneous Markov chain on infinitely (countably) many states. This framework Also allows for the possibility of recasting the model in terms of a time-\textit{inhomogeneous} Markov chain so as to allow for the evolution of the rate of evolution, but in this paper we remain in the time-homogeneous case for notational simplicity. No exotic constructs (like fragments, immortal or normal links, etc.) are used. This will maximize the number of probabilistic and mathematical tools that can be used to analyze this model. In the literature, this kind of mathematical environment is referred to as a ``Random Substitution'' \cite{Koslicki2011} as mathematicians refer to a insertions, deletions, and substitutions collectively as ``substitutions.''

Being that this model is cast in the robust, well studied environment of countable state Markov chains, exact, closed form solutions to many statistical questions become possible. Thus we can in some cases circumvent the nuanced issues introduced with estimation procedures (like the issue of attaining a local maximum versus global maximum for Hidden Markov Model MLE parameter estimation \cite{Durbin1998}, and other such issues mentioned in \cite{Miklos2009},\cite{Miklos2004}).

\section{Definition of the Model}
\label{definitions}

We present now the rigorous definition of the model. We proceeded in two steps: First we define an infinite state Markov chain to incorporate insertions and substitutions. Second we construct the induced reversible Markov chain which will incorporate deletions. To construct the first Markov chain, we need:
\begin{enumerate}
\item $\mathscr{A}=\{a_1,\dots, a_t\}$ a finite set of ordered symbols referred to as an \textit{alphabet}.
\item For each letter $a\in \mathscr{A}$, $(\Omega_a,P_a)$ a finite (non-empty) probability space.
\item For each letter $a\in \mathscr{A}$, a function $g_a:\Omega_a \rightarrow \mathscr{A}^*$ with the property that if $|g_a(\omega)|>1$, then the word $g_a(\omega)$ begins with $a$.
\end{enumerate}
Let $\mathscr{A}^n$ denote all words of length $n$ formed from letters of $\mathscr{A}$. Then $\mathscr{A}^*=\cup_{n\geq 1} \mathscr{A}^n$ is the set of finite length words formed from $\mathscr{A}$. 
%For $w=b_1\dots b_n\in \mathscr{A}^*$ and $v\in \mathscr{A}$ words, we set $|w|=n$ and let $|w|_v$ denote the number of occurrences of $v$ in $w$ with overlaps. 
%For $n\in \mathbb{N}$, let $L_n(w)=(|w|_{v_1},\dots, |w|_{v_{t_n}})^T$ be the vector of occurrences of all $n$-length words $v_i\in \hat{A}^n$ where $\hat{A}^n$ is the intersection of $A^n$ and all $n$-length words producible with non-zero probability by the Markov chain defined in the next section. We set $L(w)=L_1(w)$.\\
The alphabet $\mathscr{A}$ is usually equal to either $\{A,C,T,G\}$ for DNA models, or $\{R,H,K,\dots,V\}$ for amino acid models. The probability spaces $(\Omega_a,P_a)$ encapsulate the probabilities of insertion and substitution. In particular, the cardinality of $\Omega_a$ gives the number of different substitution, insertion, and deletion types that are allowed to occur at the letter $a\in \mathscr{A}$. The functions $g_a:\Omega_a \rightarrow \mathscr{A}^*$ (and particularly, the ranges of functions $g_a$) specify the set of allowable substitutions and insertions. In particular, if one wishes to allow the substitution of the letter $b\in \mathscr{A}$ to occur at the letter $a\in \mathscr{A}$, then the function $g_a$ should evaluate to $b$ on some $P_a$-non-zero element $\omega_1$ of $\Omega_a$: $g_a(\omega_1)=b$. If one wishes to allow the insertion of the $n$-length word $v_1\dots v_n \in \mathscr{A}^n$ to occur after the letter $a\in \mathscr{A}$, then the function $g_a$ should evaluate to $av_1\dots v_n$ on some $P_a$-non-zero element $\omega_2$ of $\Omega_a$: $g_a(\omega_2)=av_1\dots v_n$. Notice that the word $v_1\dots v_n$ is preceded by $a$ in $g_a(\omega_2)=av_1\dots v_n$, this is to assure that $v_1\dots v_n$ has been genuinely inserted into the sequence. Lack of the initial $a$ would cause the net effect of $a$ being deleted, then $v_1\dots v_n$ being inserted into the created gap.

We now define the Markov chain representing random substitutions and insertions.
\begin{definition}[Random Substitution-Insertion]\label{rand sub def}
A random substitution-insertion (RSI) (with fixed $\mathscr{A}$, $\{(\Omega_a,P_a)\}_{a\in \mathscr{A}}$, and $\{g_a\}_{a\in \mathscr{A}}$) is an infinite state Markov chain $(\mathscr{A}^*,P)$ with transition operator $P$ defined in the following way.  For $u=b_1\dots b_n\in \mathscr{A}^*$ a word, we let $\Omega_u=\Omega_{b_1}\times \dots \times \Omega_{b_n}$ and $P_u=P_{b_1}\times \dots \times P_{b_n}$. We define $g_u:\Omega_u\rightarrow \mathscr{A}^*$ via concatenation of words: for $\omega=(\omega_1,\dots,\omega_n)\in \Omega_u$, $g_u(\omega)=g_{b_1}(\omega_1)\dots g_{b_n}(\omega_n)$. Now define $P$ by \begin{align}\label{transition probability} P(u,v)=\sum_{\omega \in g_u^{-1}(v)}P_u(\omega)\end{align}
\end{definition}

So one can think of the given model in the following way: instead of modeling the evolution of individual nucleotides (or fragments) the Markov transition operator $P$ operates on entire sequences. Indeed, the state space of this model is $\mathscr{A}^*$: the set of all sequences. The transition operator $P(u,v)$ (probability of transition from the sequence $u$ to the sequence $v$) takes into consideration \textit{every} insertion and substitution possible in one time unit to compute the appropriate probability.

We now construct the induced reversible Markov chain, which will serve to incorporate deletions. Recall first the definition of a reversible Markov chain:

\begin{definition}[Reversible Markov chain]\label{reversible Markov chain}
A Markov chain $(X,P)$ with state space $X$ and transition operator $P$, is said to be reversible if there exists a function $m:X \rightarrow (0,\infty)$ such that for all $x,y\in X$,
$$m(x)P(x,y)=m(y)P(y,x)$$
\end{definition}

We now define the Markov chain which will serve as our comprehensive model of molecular evolution.

\begin{definition}[Reversible Random Substitution-Insertion-Deletion]
\label{reversible}
we define the random substitution, insertion, deletion (RSID) model $(\mathscr{A}^*,R)$ to be the Markov chain with the transition matrix given by

\begin{align}\label{R}
R(x,y)=\frac{P(x,y)+P(y,x)}{1+\sum_{z\in \mathscr{A}^*}P(z,x)}
\end{align}
\end{definition}
This Markov chain is reversible with $m(x)$ given by $m(x)=1+\sum_{z\in \mathscr{A}^*}P(z,x)$. Since any random substitution has finite range $|\{y:P(x,y)>0\}|<\infty$, the above is well defined. Note that as mentioned in the introduction, we can completely circumvent the reversibility criterion (and simultaneously allow for different rates for insertions and deletions) by modifying the above definition in the following way. If we wish insertions to have a rate $\pi_i$ and deletions to have a rate of $\pi_d$ (one can easily make these rates depend on time, or location in a given sequence), then we can use the RSID model 
$(\mathscr{A},R)$ with $R$ given by 
\begin{align}
R(x,y)=\frac{\pi_i P(x,y)+\pi_d P(y,x)}{1+\sum_{z\in \mathscr{A}^*}P(z,x)}
\end{align}

%\begin{proposition}
%A random substitution-insertion-deletion model $(\mathscr{A}^*,R)$ is a reversible Markov chain.
%\end{proposition}
%\begin{proof}
%Let
%$$m(x)=1+\sum_{z\in \mathscr{A}^*}P(z,x)$$
%Then $m(x)R(x,y)=P(x,y)+P(y,x)=m(y)R(y,x)$.
%\end{proof}
Summarizing, $(\mathscr{A}^*,R)$ is an infinite state, discrete-time Markov chain, with the state space representing entire DNA or amino acid sequences.
We have implemented this RSID model as a discrete-time Markov since we wish to model mutations that occur due to inherent DNA replication infidelities, not mutations due to environmental factors. Such mutations can be accurately modeled in a discrete time fashion.
The transition probability $R(u,v)$ between two sequences $u$ and $v$ takes into account every possible substitution that could have happened when evolving the sequence $u$ into $v$ in one evolutionary step (the time unit can be taken to be a single replication). The transition probability $R(u,v)$ also takes into account all possible substitution and insertion paths leading from $u$ to $v$, as well as all possible substitution and deletion paths leading from $u$ to $v$. Again, the first assumption is that both insertions and deletions do not simultaneously happen in one step. Rather, to allow insertions \textit{and} deletions, one needs to consider the $n$-th step transition matrix: $R^{(n)}(u,v)$ which is the $(u,v)^{\rm th}$ entry in the $n^{\rm th}$ matrix product of $R$ with itself. This $n$-step transition probability represents the probability (summed over all possible paths) of evolving the sequence $u$ into the sequence $v$ in $n$ time units. So for the purpose of measuring the total evolutionary distance from $u$ to $v$, one considers the so called \textit{Green's function}:

\begin{align}
\label{Green's function}
G(u,v)=\sum_{n=1}^\infty R^{(n)}(u,v)
\end{align}
The Green's function represents the total measure of ever evolving the sequence $u$ into the sequence $v$. The function $G(\cdot,\cdot)$ can then be used as a measure of evolutionary distance between sequences (with the obvious applications to say, phylogenetics).

\section{Example}
\label{example}

We present here a toy example to elucidate the above definitions. We also present notation that succinctly summarizes given model. In this example, say we wish to have a mathematical object that represents a model where substitutions are described by Kimura's two parameter model \cite{Kimura1980} with transition probability equal to $0.2$ and transversion probability equal to $0.1$. Thus the instantaneous rate matrix for substitutions has the form:

\begin{align*}
Q=\begin{tabular}{r} $A$\\$C$\\ $T$ \\ $G$\end{tabular}\stackrel{\begin{tabular}{cccc}$A\hspace{0ex}$ & $C\hspace{0ex}$ & $T\hspace{0ex}$ & $G$ \end{tabular}}{\left[\begin{tabular}{llll}0.6 & 0.1 & 0.1 & 0.2\\ 0.1 & 0.6 & 0.2 & 0.1\\ 0.1 & 0.2 & 0.6 & 0.1\\ 0.2 & 0.1 & 0.1 & 0.6\end{tabular} \right]}
\end{align*}

Say we also desired the model to allow only one letter insertions or deletions, with the probability of an indel occurring being $0.01$, and the choice of the particular indel type being given by the uniform distribution. Thus, transitions occur with probability $0.2*(1-0.01)=0.198$, transversions occur with probability $0.1*(1-0.01)=0.099$ and insertions and deletions both occur with probability $0.01/4=0.0025$.

Using the notation introduced in section \ref{definitions}, the alphabet is given by $\mathscr{A}=\{A,C,T,G\}$. Each of the probability spaces $\Omega_a$ consist of eight elements. We provide the details regarding $\Omega_A, P_A,$ and $g_A$ since the other definitions are completely analogous. Now as stated $\Omega_A=\{1,2,3,4,5,6,7,8\}$ since there are eight allowable events that can happen at the base $A$: substitution to one of four other bases, plus four possible indels. Also, $P_A$ and $g_A$ are defined by
\begin{center}
\begin{tabular}{l|ll}
$\Omega_A$ & $P_A$ & $g_A$\\
\hline 1 & 0.594 & $A$\\
2 & 0.099 & $C$\\
3 & 0.099 & $T$\\
4 & 0.198 & $G$\\
5 & 0.0025 & $AA$\\
6 & 0.0025 & $AC$\\
7 & 0.0025 & $AT$\\
8 & 0.0025 & $AG$\\
\end{tabular}
\end{center}
So, for example, we have that $P_A(1)=0.594$ and $g_A(1)=A$. Note that the last four rows of the above table represent the insertion or deletion of $A,C,T,$ or $G$ respectively. To succinctly represent the probability spaces and functions for this particular example, we use the following notation: let $\Sigma$ represent the $n$-th coordinate process random variable corresponding to the Markov chain given in definition \ref{reversible} whose parameter values were determined above. Then we can represent $\Sigma$ utilizing the notation given in table \ref{notation example}.

%\newpage

\begin{table}[ht!]
\caption{Notation describing the Markov chain example from section \ref{example}}
\label{notation example}
\begin{align*}
\Sigma:\left\{
\begin{tabular}{c}
$A\rightarrow$$ \left\{ \begin{tabular}{l}$A\ {\rm with\ probability\ } 0.594$\\
$C\ {\rm with\ probability\ } 0.099$\\
$T\ {\rm with\ probability\ } 0.099$\\
$G\ {\rm with\ probability\ } 0.198$\\
$AA\ {\rm with\ probability\ } 0.0025$\\
$AC\ {\rm with\ probability\ } 0.0025$\\
$AT\ {\rm with\ probability\ } 0.0025$\\
$AG\ {\rm with\ probability\ } 0.0025$\\
\end{tabular} \right.$ \\ \\ 
$C\rightarrow$$ \left\{ \begin{tabular}{l}$A\ {\rm with\ probability\ } 0.099$\\
$C\ {\rm with\ probability\ } 0.594$\\
$T\ {\rm with\ probability\ } 0.198$\\
$G\ {\rm with\ probability\ } 0.099$\\
$CA\ {\rm with\ probability\ } 0.0025$\\
$CC\ {\rm with\ probability\ } 0.0025$\\
$CT\ {\rm with\ probability\ } 0.0025$\\
$CG\ {\rm with\ probability\ } 0.0025$\\
\end{tabular} \right.$ \\ \\ 
$T\rightarrow$$ \left\{ \begin{tabular}{l}$A\ {\rm with\ probability\ } 0.099$\\
$C\ {\rm with\ probability\ } 0.198$\\
$T\ {\rm with\ probability\ } 0.594$\\
$G\ {\rm with\ probability\ } 0.099$\\
$TA\ {\rm with\ probability\ } 0.0025$\\
$TC\ {\rm with\ probability\ } 0.0025$\\
$TT\ {\rm with\ probability\ } 0.0025$\\
$TG\ {\rm with\ probability\ } 0.0025$\\
\end{tabular} \right.$ \\ \\ 
$G\rightarrow$$ \left\{ \begin{tabular}{l}$A\ {\rm with\ probability\ } 0.198$\\
$C\ {\rm with\ probability\ } 0.099$\\
$T\ {\rm with\ probability\ } 0.099$\\
$G\ {\rm with\ probability\ } 0.594$\\
$GA\ {\rm with\ probability\ } 0.0025$\\
$GC\ {\rm with\ probability\ } 0.0025$\\
$GT\ {\rm with\ probability\ } 0.0025$\\
$GG\ {\rm with\ probability\ } 0.0025$\\
\end{tabular} \right.$ \\ \\ \end{tabular}\right.
\end{align*}
\end{table}

\section{Flexibility of the Model}
In this section we present the flexibility of the RSID model given in definition \ref{reversible Markov chain}.
\paragraph{Traditional Substitution Models of Molecular Evolution}
The RSID model  allows for the implementation of most previous substitution models of molecular evolution. For example, by using the alphabet $\mathscr{A}=\{A,C,T,G\}$, and for each $a\in \mathscr{A}$, letting $\Omega_a=\{1,2,3,4\}$, and $g_a(\Omega_a)=\{A,C,T,G\}$, and choosing the probabilities $P_a$ appropriately, the RSID model given in definition \ref{reversible} completely encompasses the JC \cite{Jukes1969}, HKY \cite{Hasegawa1985}, FEL81 \cite{Felsenstein1981}, K2P \cite{Kimura1980}, and REV \cite{Tavare1986} models. In fact, in this particular case the RSID model is a generalization of \textit{all} possible homogeneous rate Markov models of DNA or amino acid evolution. This is due to the fact that if $g_a(\Omega_a)=\mathscr{A}$ for each $a\in \mathscr{A}$, then the RSID model simply becomes a traditional finite state Markov chain (with as many states as the cardinality of the alphabet $\mathscr{A}$).

\paragraph{Heterogeneous rates}
Models utilizing heterogeneous rates of evolution can be introduced by slightly modifying definition \ref{rand sub def} and consequently \ref{reversible Markov chain}. Instead of fixed probability spaces $(\Omega_a,P_a)$, we allow the probability space to change. Let $\mathscr{P}(g_a(\Omega_A))$ denote the set of probability measures on $g_a(\Omega_a)$. Then the desired heterogeneity can be introduced with a \textit{random probability} (also known as a random element in the literature \cite{Billingsley1968}) i.e. a probability-valued random variable:  $X_a:(\Omega_u,P_u)\rightarrow \mathscr{P}(g_a(\Omega_a))$. Then definitions \ref{rand sub def} and \ref{reversible} work just as before, but instead by utilizing the spaces $(\Omega_a,X_a)$. Hence, the RSID model can also incorporate heterogeneous evolution rates. This is similar in spirit to the idea of the ``variety of fragments'' utilized in \cite{Thorne1992}.

\paragraph{Neighboring Dependencies}
We can introduce neighboring dependencies by again slightly modifying definition \ref{rand sub def}. For $u=b_1\dots b_n\in \mathscr{A}^*$, instead of using the probability $P_u=P_{b_1}\times \dots \times P_{b_n}$, we can use a different probability $P_u$ whose marginal distributions correspond to the $P_{b_1},\dots,P_{b_n}$. This is simply a utilization of a \textit{coupling}. A coupling between two probability spaces $(\Omega_1,P_1)$ and $(\Omega_2,P_2)$ is a probability $C$ on the space $\Omega_1\times \Omega_2$ whose marginal distributions are $P_1$ and $P_2$. Hence the original definition \ref{rand sub def}, which assumes that what happens at a specific nucleotide (be it substitution, insertion, or deletion) is independent of its neighbors, simply uses the null coupling. Of course, the specific coupling to be used depends on the situation at hand, we are simply elucidating the various mathematical constructs that may be used to implement the desired biological properties.

\paragraph{Parameterization}
Now the RSID model is not meant to be implemented in its most general form, but rather parameterized to a certain degree taking into consideration the problem at hand. For example, we sketch here a possibility of parameterizing the indel appearance rate to depend only on a two parameter $\alpha, \beta$ power law $\alpha L^{-\beta}$ on the length $L$ of the particular indel. To accomplish this, all one needs to do is define the $P_a$ in the following way: for $\omega \in \Omega_a$, let $P_a(\omega)=\alpha |g_a(\omega)|^{-\beta}$. Such an RSID would then be able to accurately model, for example, the human ribosomal protein pseudogene evolution as studies in \cite{Zhang2003}. Further nuanced parameterizations are possible and easily implemented into the RSID model. For example, one can easily use a distribution on the possible indels that not only takes length into consideration, but also GC content.

\paragraph{Implementable Biological Phenomena}
Due to the flexibility of this model, we summarize here the various biological phenomena that can be implemented by using the RSID model and its variants mentioned above. The RSID model provides a systematic framework for studying models of molecular evolution that implement heterogeneous rates, conservation of reading frame (through careful selection of the functions $g_a$), variation in conservation, differing rates of insertion and deletion, customizable parameterization of the probabilities and types of substitutions, insertions, and deletions available (through the specification of the probabilities $(\Omega_a,P_a)$), as well as neighboring dependencies.

\section{Algorithms and Implementation}
The aim of this paper is not to present a preprogrammed set of algorithms, but rather a comprehensive framework to be adapted to the specific situation at hand. So instead of enumerating a plethora of algorithms that can be used in a variety of implementations, we rather present a general method for producing efficient algorithms useful for computation. We later will also show how setting the model on a firm mathematical footing allows for the possibility for general theorems to be applied. This will present new avenues for algorithmic implementations.

One of the computational difficulties is in the calculation of the transition probabilities (see equation \ref{transition probability}). Rewriting equation \ref{transition probability}, we obtain
\begin{align}
\label{expanded transition probability}
P(u,v)=\sum_{\omega \in g_u^{-1}(v)} P_{u_1}(\omega_1)P_{u_2}(\omega_2)\dots P_{u_n}(\omega_n)
\end{align}

Now the above equation \ref{expanded transition probability} is an example of the so called \textit{sum-product} algorithm: since the underlying factor graph is a tree, the results of Kschischange et al. \cite{Kschischang2001} can be applied to develop an algorithm that calculates this sum in linear time. Hence the full RSID transition probability in equation \ref{R} can also be calculated in linear time. Dynamical programming methods can also be used at this step.

Now if one wishes to measure the total (evolutionary) distance between the sequences $u$ and $v$ by using the Green's function $G(u,v)$ found in equation \ref{Green's function}, it is typically intractable to attempt to compute the entire infinite sum $\sum_{i=1}^{\infty} R^{(i)}(u,v)$, but in \cite{Koslicki2011}, it is proved that the approximation
\begin{align}
\label{Green's function approximation}
G_n(u,v)=\sum_{i=1}^{n} R^{(i)}(u,v)
\end{align}
converges to the full summation $G(u,v)$ exponentially quickly. So in practice one needs only calculate $G_n(u,v)$ for some adequately large value of $n$ (representing one to $n$ generations of mutations from $u$ to $v$). Then \cite{Koslicki2011} provides error estimates for this approximation. Now $G_n(\cdot,\cdot)$ is an infinite matrix, but the entire basis of $\mathscr{A}^*$ need not be used. In fact, at worst, in the computation of $G_n(u,v)$ one needs only the basis elements $\cup_i^m \mathscr{A}^i$ for $m=|u|\left(\max_{a\in \mathscr{A}\\ \omega \in \Omega_a}|g_a(\omega)|\right)^n$. In practice, this large of a basis is not needed and one can utilize far fewer elements. Since DNA sequences are not free to explore the entire space $\mathscr{A}^*$ one can use, for example, GC-content or entropy requirements to eliminate extraneous basis elements. After obtaining an appropriate basis, the matrix sum $G_n(u,v)$ can be computed using, for example, the $\mathscr{O}(n\log n)$ algorithm found in \cite{Hu1982} and \cite{Hu1984}.

Again, since this article is meant to introduce an analytic environment in which to study biologically accurate models of molecular evolution, we do not provide explicit algorithms, but have rather indicated that such algorithms can be developed and tailored to the specific problem at hand. Being that the transition probabilities can be calculated in linear time, and then Green's function in loglinear time, the main computational issue is determining an appropriate basis of ``biologically viable'' sequences for the computation of $G_n$. This basis determination can be accomplished by a combination of various techniques including GC-content and entropy methods.

\paragraph{Alternative Algorithms}
We now demonstrate the advantage of having a well-developed mathematical background underlying the RSID model by presenting an alternative method of implementing the calculation of the Green's function $G(u,v)$. Utilizing a slight modification of the proof of the spectral theorem for self-adjoint compact operators (\cite{Dunford1988}, Theorem 1, pg. 895), it can be shown that $R$ is diagonalizable. Utilizing this diagonalizability, on can compute $R^{(n)}$ and hence $G$ in a straightforward theoretical fashion. This provides another avenue for investigating the possible algorithmic implementations of the RSID model.

\section{Parameter Estimation}

Once a particular parameterization choice has been made, the problem of parameter estimation can be approached in a number of ways. While parameters can be estimated using MLE methods via counting indel and substitution types and frequencies over a number of optimal and sub-optimal alignments (or all alignments), we present here a new \textit{alignment free} method of parameter estimation.
%free from some of the convergence issues that hamper other MLE methods (for example, finding a local instead of a global maximum likelihood).
The method is based on frequencies of occurrence of subwords and follows the exposition of \cite{Koslicki2011}. The main argument is based on the fact that while one may assume the independence of substitution and indel \textit{events} that occur at neighboring nucleotides, this does not in general imply the independence of the frequency of occurrence of neighboring \textit{nucleotides} (or $n$-mers of nucleotides) due to the inclusion of indels. This fact can be taken advantage of to estimate parameters. Briefly, the frequencies can be calculated explicitly (as in \cite{Koslicki2011}), and then these are used to assist in a maximum likelihood estimation. We begin by sketching how one can calculate the expected frequency of $n$-mers.

First we need a vector of occurrences: For $w\in \mathscr{A}^*$, $a_i\in \mathscr{A}$, let $F_1(w)=(|w|_{a_1},\dots, |w|_{a_n})$ denote the column vector of occurrences of single letters in the sequence $w$. Let also $\Sigma_m$ denote the $n^{\rm th}$ coordinate process associated to $(R,\mathscr{A}^*)$ where letters are taken $m$ at a time with overlaps (detail in \cite{Koslicki2011}. We also define the matrices
\begin{definition}[Substitution Matrices]
For a given RSID Markov chain $(R,\mathscr{A}^*)$ and integers $m,n$, define the $m$-th substitution matrix entry wise as follows: for $i,j\in \mathscr{A}^*$,
\begin{align}
\label{substitution matrix}
\left(M^{(m)}_{\Sigma_n} \right)_{ij}=\mathbb{E}_j|\Sigma^{(m)}_n|_i
\end{align}
\end{definition}
Here $|w|_i$ means the number of occurrences (with overlap) of $i$ in $w$, and the $\mathbb{E}_j$ denotes expectation taken with unit mass on the Markov state corresponding to $j$. Let $\Lambda$ denote the dominant eigenvalue of $M_{\Sigma_1}^{(1)}$. The following was proved in \cite{Koslicki2011}
\begin{lemma}\label{occurrence convergence}
For an RSID Markov chain $(R, \mathscr{A}^*)$, the sequence of vectors $\frac{F_1(\Sigma_n(w))}{\Lambda^n}$ converges almost surely and exponentially quickly to the strictly positive eigenvector of $M^{(1)}_{\Sigma_1}$ corresponding to $\Lambda$ with this eigenvector being independent of $w\in \mathscr{A}^*$.
\end{lemma}
Normalizing the strictly positive eigenvector of $M^{(1)}_{\Sigma_1}$ corresponding to $\Lambda$ gives the expected frequency of appearance of the letter $\mathscr{A}$. This implies that the observation frequencies converge exponentially quickly to the expected frequencies. Similar theorems in \cite{Koslicki2011} are proved for arbitrary $n$-mers of letters from $\mathscr{A}$. These frequencies are given in terms of the various parameters that were chosen for the probabilities $P_a$ in the definition of \ref{rand sub def}. For single letters frequencies in the DNA case, this results in four constraints (linear in the entires of $M_{\Sigma_1}^{(1)}$), and hence can estimate up to four parameters. Counting couplets of nucleotides gives 16 quadratic constraints; in general, counting $n$-mers gives $4^n$ equations of degree $n$ in the entires of $M_{\Sigma_1}^{(1)}$. Setting these equations equal to the observed frequencies, one can then use standard nonlinear optimization techniques.

For example, if one was given a data set and wished to model it using an RSID that used the Kimura two parameter $\gamma,\delta$ model (\cite{Kimura1980}) to describe the substitutions and a two parameter power law ($P_a(\omega)=\alpha |g_a(\omega)|^{-\beta}$) to describe the indel distribution, one would only need to count single letter frequencies to obtain four equations in the four parameters $\alpha, \beta, \gamma, \delta$. The maximal likelihood parameter estimator can then be found using any preferred optimization technique.

It is important to note that this method of parameter estimation is completely alignment free. This circumvents the myriad issues involved when, for example, estimating parameters in a classical substitution model of molecular evolution: choosing a particular alignment algorithm, a particular alignment parameterization (linear, log-linear, affine), particular mismatch, match, gap opening, and gap extensions penalties. It is hard to overstate the advantages of having an alignment-free parameterization technique. This is mainly due to the fact that choosing a particular alignment scheme is a nuanced endeavor where slight changes in implementation can lead to large changes in alignment outcome. Furthermore, it has been observed that various algorithms have potential to introduce hidden bias (see \cite{Lunter2008}, \cite{Lunter2007}, \cite{Metzler2003}, \cite{Metzler2001}, \cite{Fleissner2000}, \cite{Vingron1994}, \cite{Fitch1983}).

\section{Application to Human Ribosomal Protein Pseudogenes}

We applied a simple implementation of the RSID model to the data set used in \cite{Zhang2003} to measure the indel length distribution in human ribosomal protein pseudogenes. Note that we wish to measure the underlying intrinsic indel length distribution in the evolution of human ribosomal protein pseudogenes, not (as \cite{Zhang2003} did) to estimate this distribution via comparison with a different species. We used the Kimura two-parameter model (\cite{Kimura1980}) for the substitutions and a two parameter power law ($P_a(\omega)=\alpha |g_a(\omega)|^{-\beta}$) to describe the indel distribution. We also assumed reversibility and time homogeneity. Using the parameter estimation technique detailed above, we obtained a power law of indel length $L$ distribution of $.4955 L^{-1.4040}$. Compare this to the result of \cite{Zhang2003} which gave a deletion distribution of $.48 L^{-1.51}$ and insertion distribution of $.53 L^{-1.60}$. The discrepancy is most likely due to the fact that we considered insertions and deletions together as well as assumed reversibility, whereas the results of \cite{Zhang2003} clearly indicate different rates for insertions and deletions. However, the basic power-law similarity indicates the feasibility of our alignment free parameter estimation technique.

\section{Conclusion}

We have presented a comprehensive new framework for handling biologically accurate models of molecular evolution. As we have demonstrated, the number of implementable biological phenomenon is vast. One profound advantage of stating the RSID in the language of an infinite state Markov chain is that one can utilize the vast mathematical literature to rigorously analyze a given implementation. We used such theorems to develop an alignment-free parameter estimation technique. This alignment-free parameter estimation technique circumvents many nuanced issues related to alignment-dependent estimation. 

We then applied one possible implementation of the RSID model to analyze the power-law governing the indel distribution in human ribosomal protein pseudogene sequences \cite{Zhang2003}. Our analysis corroborated the observations of \cite{Zhang2003}. It is of great interest to note that when it comes to indel length distribution, our RSID model gave similar power law parameters as \cite{Zhang2003} without needing to consider the complicated alignments described in the ``Sequence Alignment'' section of \cite{Zhang2003}.


\begin{thebibliography}{10}

\bibitem{Billingsley1968}
P.~Billingsley.
\newblock {\em {Convergence of probability measures}}.
\newblock John Wiley \& Sons, New York, NY., 1968.

\bibitem{Bradley2007}
R.~K. Bradley and I.~Holmes.
\newblock {Transducers: an emerging probabilistic framework for modeling indels
  on trees.}
\newblock {\em Bioinformatics (Oxford, England)}, 23(23):3258--62, Dec. 2007.

\bibitem{Cartwright2009}
R.~Cartwright.
\newblock {Problems and solutions for estimating indel rates and length
  distributions.}
\newblock {\em Molecular biology and evolution}, 26(2):473--80, Feb. 2009.

\bibitem{Dunford1988}
N.~Dunford and J.~T. Schwartz.
\newblock {\em {Linear operators part II spectral theory}}.
\newblock Wiley-Interscience, 1988.

\bibitem{Durbin1998}
R.~Durbin.
\newblock {\em {Biological sequence analysis: probabilistic models of proteins
  and nucleic acids}}.
\newblock Cambridge Univ Pr, 1998.

\bibitem{Felsenstein1981}
J.~Felsenstein.
\newblock {Evolutionary trees from DNA sequences: A maximum likelihood
  approach}.
\newblock {\em Journal of Molecular Evolution}, 17:368--376, 1981.

\bibitem{Fitch1983}
W.~M. Fitch and T.~F. Smith.
\newblock {Optimal sequence alignments.}
\newblock {\em Proceedings of the National Academy of Sciences of the United
  States of America}, 80(5):1382--1386, Mar. 1983.

\bibitem{Fleissner2000}
R.~Flei\ss ner, D.~Metzler, and A.~von Haeseler.
\newblock {Can one estimate distances from pairwise sequence alignments?}
\newblock In E.~Bornberg-Bauer, U.~Rost, J.~Soye, and M.~Vingron, editors, {\em
  Proceedings of the German conference on bioinformatics}, pages 89--95,
  Heidelberg, Berlin, 2000. Logos.

\bibitem{Hasegawa1985}
M.~Hasegawa, H.~Kishino, and T.~Yano.
\newblock {Dating of the human-ape splitting by a molecular clock of
  mitochondrial DNA}.
\newblock {\em Journal of molecular evolution}, 22(2):160--174, 1985.

\bibitem{Hu1982}
T.~Hu and M.~Shing.
\newblock {Computation of matrix chain products part I}.
\newblock {\em SIAM J. Comput.}, 11(2):362--373, 1982.

\bibitem{Hu1984}
T.~Hu and M.~Shing.
\newblock {Computation of matrix chain products part II}.
\newblock {\em SIAM J. Comput.}, 13(2):228--251, Feb. 1984.

\bibitem{Jukes1969}
T.~Jukes and C.~Cantor.
\newblock {\em {Evolution of protein molecules}}, pages 21--132.
\newblock Academic Press, New York, NY., 1969.

\bibitem{Kim2007}
J.~Kim and S.~Sinha.
\newblock {Indelign: a probabilistic framework for annotation of insertions and
  deletions in a multiple alignment.}
\newblock {\em Bioinformatics (Oxford, England)}, 23(3):289--97, Feb. 2007.

\bibitem{Kimura1980}
M.~Kimura.
\newblock {A simple method for estimating evolutionary rates of base
  substitutions through comparative studies of nucleotide sequences}.
\newblock {\em Journal of Molecular Evolution}, 6:111--120, 1980.

\bibitem{Koslicki2011}
D.~Koslicki.
\newblock {\em {Random Substitutions, Markov Chains, and Martin Boundaries}}.
\newblock PhD thesis, Pennsylvania State University, 2011.

\bibitem{Kschischang2001}
F.~Kschischang, B.~Frey, and H.-a. Loeliger.
\newblock {Factor graphs and the sum-product algorithm}.
\newblock {\em IEEE Transactions on Information Theory}, 47(2):498--519, 2001.

\bibitem{Lio1998}
P.~Lio and N.~Goldman.
\newblock {Models of molecular evolution and phylogeny}.
\newblock {\em Genome research}, 8(12):1233--1244, 1998.

\bibitem{Lunter2007}
G.~Lunter.
\newblock {Probabilistic whole-genome alignments reveal high indel rates in the
  human and mouse genomes.}
\newblock {\em Bioinformatics (Oxford, England)}, 23(13):i289--296, July 2007.

\bibitem{Lunter2008}
G.~Lunter, A.~Rocco, N.~Mimouni, A.~Heger, A.~Caldeira, and J.~Hein.
\newblock {Uncertainty in homology inferences: assessing and improving genomic
  sequence alignment.}
\newblock {\em Genome research}, 18(2):298--309, Feb. 2008.

\bibitem{Metzler2003}
D.~Metzler.
\newblock {Statistical alignment based on fragment insertion and deletion
  models}.
\newblock {\em Bioinformatics}, 19(4):490--499, Mar. 2003.

\bibitem{Metzler2001}
D.~Metzler, R.~Fleissner, a.~Wakolbinger, and a.~von Haeseler.
\newblock {Assessing variability by joint sampling of alignments and mutation
  rates.}
\newblock {\em Journal of molecular evolution}, 53(6):660--9, Dec. 2001.

\bibitem{Miklos2004}
I.~Mikl\'{o}s, G.~Lunter, and I.~Holmes.
\newblock {A ``long indel" model for evolutionary sequence alignment}.
\newblock {\em Molecular Biology and Evolution}, 21(3):529, 2004.

\bibitem{Miklos2009}
I.~Mikl\'{o}s, A.~Nov\'{a}k, R.~Satija, R.~Lyngs\o, and J.~Hein.
\newblock {Stochastic models of sequence evolution including insertion-deletion
  events.}
\newblock {\em Statistical methods in medical research}, 18(5):453--85, Oct.
  2009.

\bibitem{Tavare1986}
S.~Tavar\'{e}.
\newblock {Some probabilistic and statistical problems in the analysis of DNA
  sequences}.
\newblock {\em Some mathematical questions in biology—DNA sequence analysis},
  17:57--86, 1986.

\bibitem{Thorne1992}
J.~Thorne and H.~Kishino.
\newblock {Freeing phylogenies from artifacts of alignment.}
\newblock {\em Molecular Biology and Evolution}, 9(6):1148--1162, 1992.

\bibitem{Thorne1991}
J.~Thorne, H.~Kishino, and J.~Felsenstein.
\newblock {An evolutionary model for maximum likelihood alignment of DNA
  sequences}.
\newblock {\em Journal of Molecular Evolution}, 33(2):114--124, 1991.

\bibitem{Vingron1994}
M.~Vingron and M.~S. Waterman.
\newblock {Sequence alignment and penalty choice. Review of concepts, case
  studies and implications.}
\newblock {\em Journal of molecular biology}, 235(1):1--12, Jan. 1994.

\bibitem{Whelan2001}
S.~Whelan, P.~Lin, and N.~Goldman.
\newblock {Molecular phylogenetics: state-of-the-art methods for looking into
  the past}.
\newblock {\em TRENDS in Genetics}, 17(5):262--272, 2001.

\bibitem{Zhang2003}
Z.~Zhang and M.~Gerstein.
\newblock {Patterns of nucleotide substitution, insertion and deletion in the
  human genome inferred from pseudogenes}.
\newblock {\em Nucleic acids research}, 31(18):5338--5348, 2003.

\end{thebibliography}
\end{document}